%% file: main.tex
\documentclass[conference,compsoc]{IEEEtran}
\IEEEoverridecommandlockouts

\ifCLASSOPTIONcompsoc
  \usepackage[nocompress]{cite}
\else
  \usepackage{cite}
\fi

\usepackage{amsmath}
\usepackage{amssymb}
\usepackage{hyperref}
\usepackage{dsfont}
\usepackage{tcolorbox}
\tcbuselibrary{theorems}
\usepackage{enumitem}
\usepackage{amsmath,amssymb,amsfonts}
\usepackage{amsthm}
\usepackage{algorithm2e}
\usepackage{multicol}
\usepackage{multirow}
\usepackage{tabularx}
\usepackage{rotating}
\usepackage{url}
\usepackage{siunitx}
\usepackage{ragged2e}
\usepackage{array,booktabs,makecell}

\newcolumntype{P}[1]{>{\RaggedRight\arraybackslash}p{#1}}

\ifCLASSOPTIONcompsoc
    \usepackage[caption=false, font=normalsize, labelfont=sf, textfont=sf,subrefformat=parens,labelformat=parens]{subfig}
\else
\usepackage[caption=false, font=footnotesize,subrefformat=parens,labelformat=parens]{subfig}
\fi

\newtcbtheorem[]{mytheo}{Application}%
{colback=green!5,colframe=green!35!black,fonttitle=\bfseries}{th}

\newtcbtheorem[]{mytheo2}{Open Question}%
{colback=blue!5,colframe=blue!35!black,fonttitle=\bfseries}{th}

\begin{document}

\title{SoK: Fighting Counterfeits with Cyber--Physical Synergy Based on \\ Physically-Unclonable Identifiers of Paper Surface}

\author{\IEEEauthorblockN{Anirudh Nakra}
\IEEEauthorblockA{University of Maryland, College Park\\
anakra@umd.edu}
\and
\IEEEauthorblockN{Min Wu}
\IEEEauthorblockA{University of Maryland, College Park\\
minwu@umd.edu}
\and
\IEEEauthorblockN{Chau-Wai Wong}
\IEEEauthorblockA{NC State University\\
chauwai.wong@ncsu.edu}}

\maketitle

\thispagestyle{plain}
\pagestyle{plain}

\begin{abstract}

\input{Sections/abstract}

\end{abstract}

\IEEEpeerreviewmaketitle

\input{Sections/sec1_introduction}

\input{Sections/sec2_cyberphysical}

\input{Sections/sec3_physicalpaperPUF}

\input{Sections/sec4_cyberframework}

\input{Sections/sec5_taxonomy}

\input{Sections/sec6_sysdesign}

\input{Sections/sec7_practical_sys_design}

\input{Sections/sec8_common_sysdesign}

\input{Sections/sec9_discussion}

\input{Sections/acknowledgments}

\bibliographystyle{IEEEtran}

\bibliography{references}

\end{document}

%% file: Sections/abstract.tex
Counterfeit products cause severe harm to public safety and health by penetrating untrusted supply chains. Numerous anti-counterfeiting techniques have been proposed, among which the use of inherent, unclonable irregularities of paper surfaces has shown considerable potential as a high-performance economical solution. Prior works do not consider supply chains cohesively, either focusing on creating or improving unclonable identifiers or on securing digital records of products. This work aims to systematically unify these two separate but connected research areas by comprehensively analyzing the needs of supply chains. We construct a generalized paper-based authentication framework and identify important shortcomings and promising ideas in the existing literature. Next, we do a stage-wise security analysis of our consolidated framework by drawing inspiration from works in signal processing, cryptography, and biometric systems. Finally, we examine key representative scenarios that illustrate the range of practical and technical challenges in real-world supply chains, and we outline the best practices to guide future research.

%% file: Sections/sec1_introduction.tex
\section{Introduction\label{Sec1}}

Counterfeiting is prevalent across different industries, including fashion, electronics, and healthcare \cite{counterfeitcbp}. These counterfeit goods reach consumers through intricate logistical distribution channels known as supply chains. Counterfeiters frequently penetrate existing untrusted supply chains, such as those of e-commerce websites  \cite{counterfeitecommerce}, to introduce counterfeit products into the market, posing significant risks to public safety and health. Counterfeit medications, which refer to recycled, expired, or fake medicines, were the most seized U.S. health and safety products in 2022 \cite{counterfeitcbp}. Alternatively, skilled forgers duplicate important identification documents such as passports and licenses, which poses a substantial threat to national security \cite{docfraud}.   

Researchers have proposed several types of anti-counterfeiting solutions in the past. Special inks \cite{cox2002digital}, laser engravings \cite{FEI2016657}, copy-resistant patterns \cite{picard2004digital}, and optically variable features \cite{ren2020optical} have been developed to secure products against possible forgery. However, such methods add a considerable overhead to the verification pipeline and often require specialized techniques in the verification process which restricts the ability of supply chain stakeholders to participate. In contrast to the class of \textit{extrinsic} signatures that add distinguishing patterns to different products, there is a growing body of literature using the \textit{intrinsic} fingerprint of hosting material \cite{sharma2011paperspeckle,toreini2017texture,beekhof2008secure,chenandzeng} to validate its authenticity. Paper-based authentication systems \cite{clarksonpuf,wongtifs,liu2018enhanced} are an economical solution that leverages the microscopic imperfections in the surface texture of the paper to authenticate products. 

Even though significant research and development efforts have been carried out to design algorithms for such systems, there has been little systematic study on their similarities, trade-offs, and security. This paper aims to enhance the understanding of these systems in a systematic, integrated way by considering the interaction of different aspects of the supply chain illustrated in Fig. \subref*{fig:overall}. In this work, we decompose the operation of the supply chain into two worlds: the physical/real world where the product exists, versus the cyber world that contains the infrastructure responsible for maintaining the products' records digitally. Prior works either focus on designing new techniques for the physical or cyber world but do not consider both worlds holistically. We introduce the two worlds, their constituents, and the challenges associated with them in Section~\ref{Sec2}. After analyzing these worlds, we identify and explain why paper surfaces have the potential to develop a secure physical world in Section~\ref{Sec3}. Next, we introduce a general framework in Section~\ref{Sec4} used to evaluate and categorize the existing anti-counterfeiting systems into a taxonomy in Section~\ref{Sec5}. Using our gained understanding, we examine the vulnerabilities of our framework, explore practical supply chains of different scales, explain possible threats, discuss attack mitigation strategies, and highlight important design considerations in Sections~\ref{Sec6}--\ref{Sec8}. Section~\ref{Sec9} concludes our survey and analysis. Specifically, our main contributions are:

\begin{figure*} 
    \centering
        \subfloat[\label{fig:overall}]{%
        \includegraphics[width=0.57\linewidth]{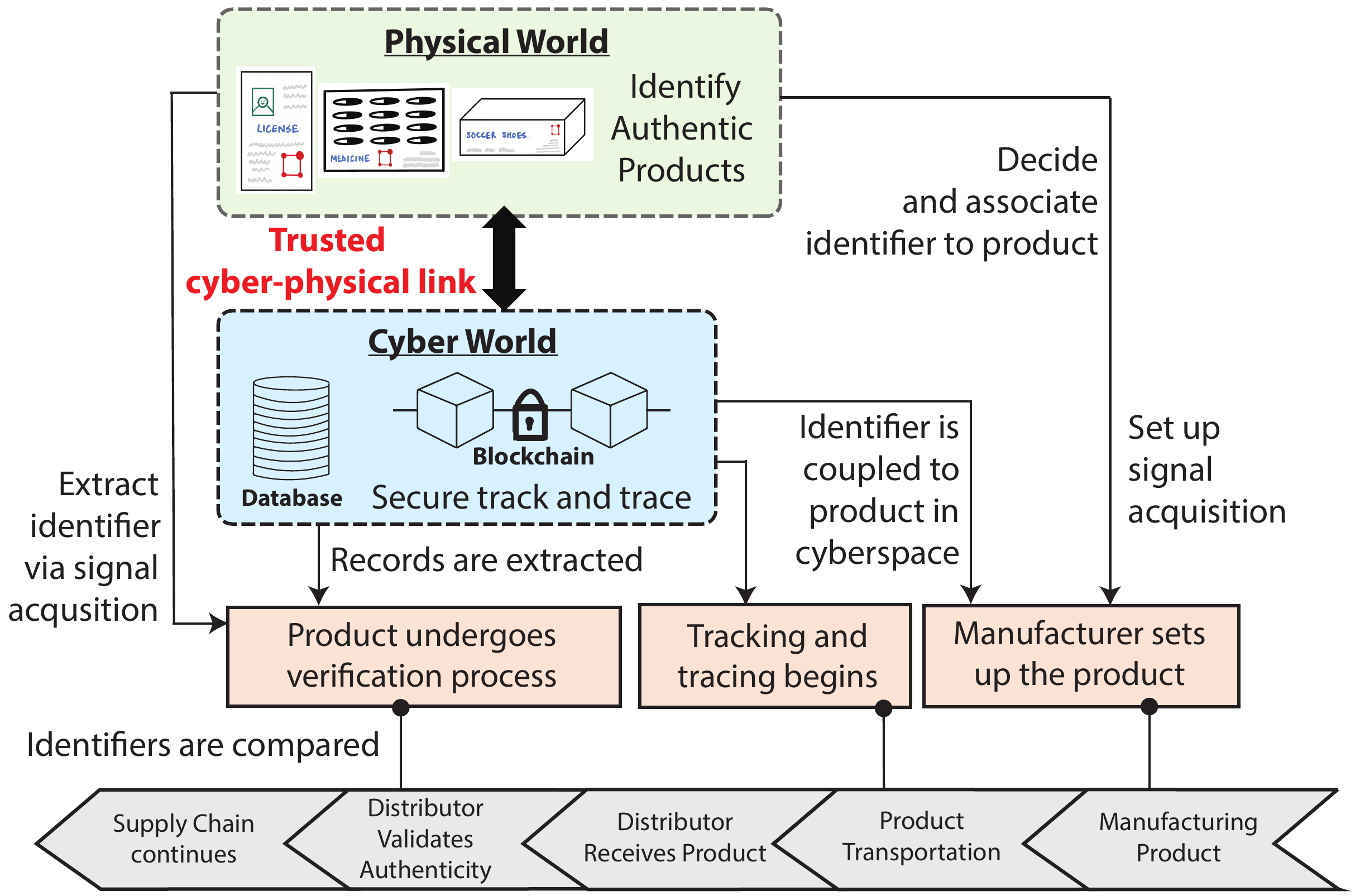}}
    \hfill
        \subfloat[\label{fig:venn}]{
       \includegraphics[width=0.4\linewidth]{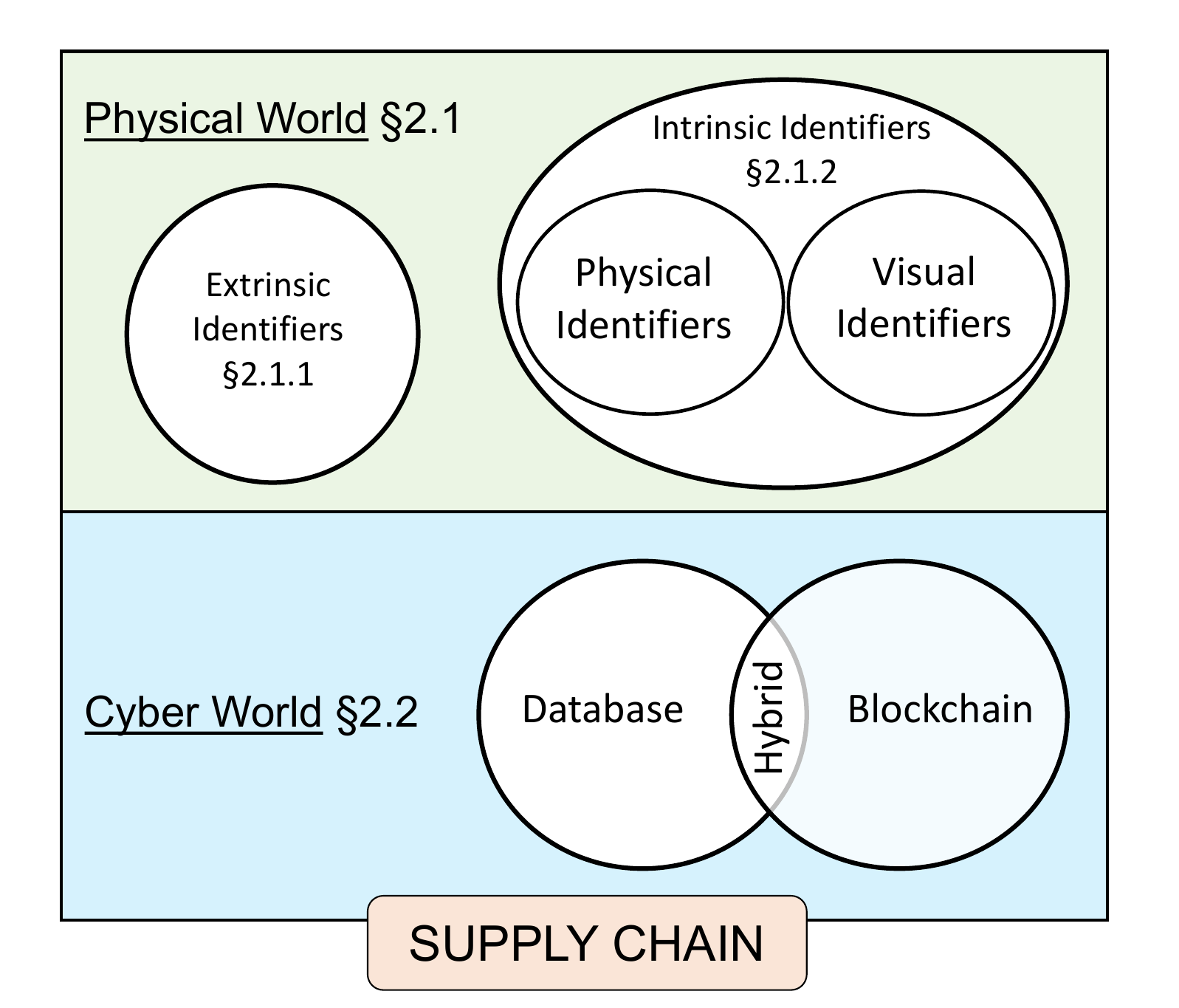}}%
  \caption{(a) An example of the dynamics of a supply chain and its constituents. (b) Demarcation of the supply chain into the physical and cyber world.}
  \label{fig1} 
\end{figure*}

\noindent

\begin{itemize}
    
    \item \textbf{Systematization of Knowledge}: We distinguish two different research aspects in combating counterfeiting: methods focusing on creating unduplicable identifiers and methods securing virtual records using distributed ledgers. We synergize these two aspects by understanding their interactions, systematizing the knowledge from the two research aspects, and integrating them into a comprehensive framework.
    
    \item \textbf{Classification and analysis of anti-counterfeiting methods}: We use our designed framework to create an extensive taxonomy and outline crucial security vulnerabilities overlooked in the existing literature.
    
    \item \textbf{Exploring practical systems}: Using our framework's security analysis, we explore different practical system designs representing a spectrum of real-life supply chain scenarios.   
    
\end{itemize}

%% file: Sections/sec2_cyberphysical.tex
\section{Dissecting Supply Chains into Two Worlds\label{Sec2}}

A supply chain is defined as the set of firms and processes that are involved in the life cycle of a product \cite{mentzer2001defining}. Due to their complexity, supply chain operations such as inventory management rely on tracking and tracing the product accurately. However, track-and-trace methods do not exist in isolation. Developing secure and robust methods involves understanding the interactions between different aspects of the supply chain. For this reason, we demarcate the supply chain into two distinct worlds as shown in Fig.~\subref*{fig:venn}. First, the real product, along with its unique physical signature, resides in the \textit{physical world}. Second, the records detailing product movements are housed in a virtual environment known as the \textit{cyber world}, which encompasses the digital infrastructure essential for tracking and tracing products. Since the supply chain operates across these two dissimilar worlds that interact with each other regularly as illustrated in Fig.~\subref*{fig:overall}), a holistic understanding of the security of both these worlds is essential. By linking an authentic product with a secure virtual record via a \textit{trusted cyber--physical link}, the authenticity of a product delivered via a supply chain can be guaranteed. Next, we examine the two worlds and their associated challenges.

\subsection{The Physical World}

\subsubsection{Extrinsic Identifiers}  

\begin{figure*}
    \centering
  \subfloat[\label{fig:cdp}]{%
       \includegraphics[width=0.15\linewidth]{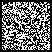}}
    \hfill
    \subfloat[\label{fig:bubbletag}]{
        \includegraphics[width=0.2\linewidth]{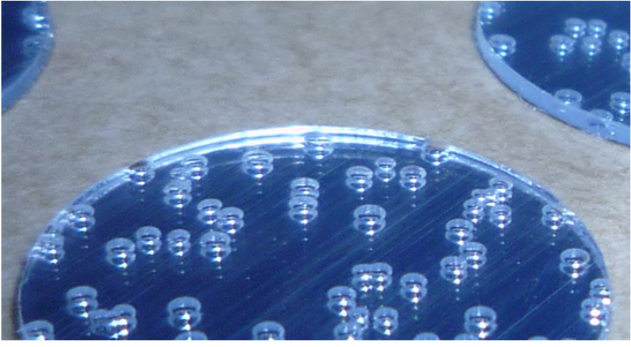}}
    \hfill
    \subfloat[\label{fig:fibertag}]{%
    \includegraphics[width=0.15\linewidth]{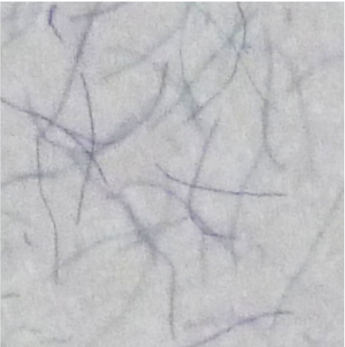}}
    \hfill
    \subfloat[\label{fig:sticker}]{%
        \includegraphics[width=0.15\linewidth]{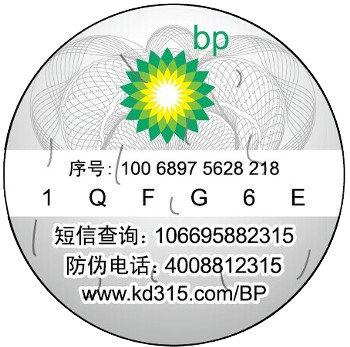}}  

    \subfloat[\label{fig:paper}]{%
    \includegraphics[width=0.18\linewidth]{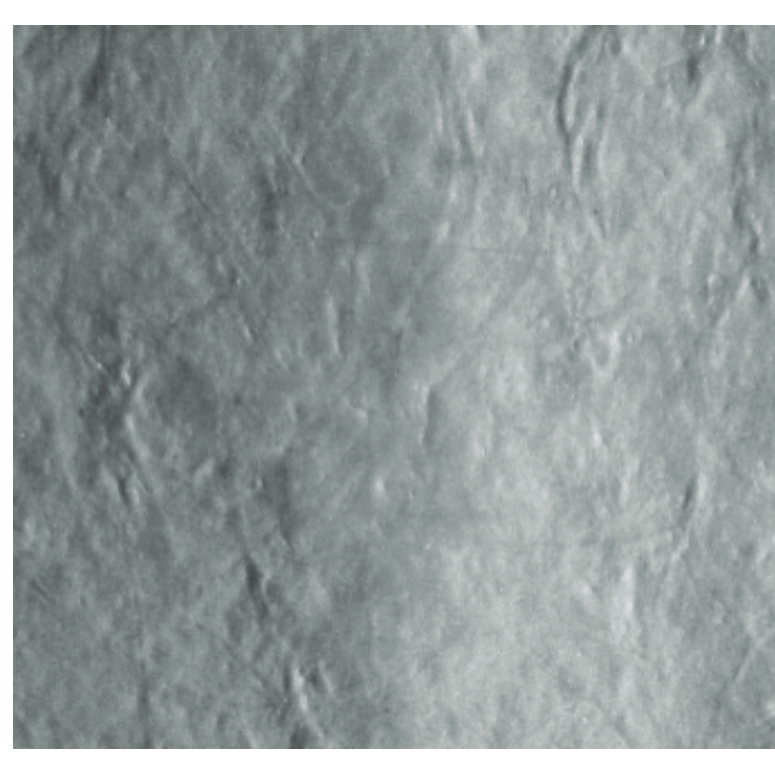}}
    \hfill
    \subfloat[\label{fig:cfm}]{%
        \includegraphics[width=0.2\linewidth]{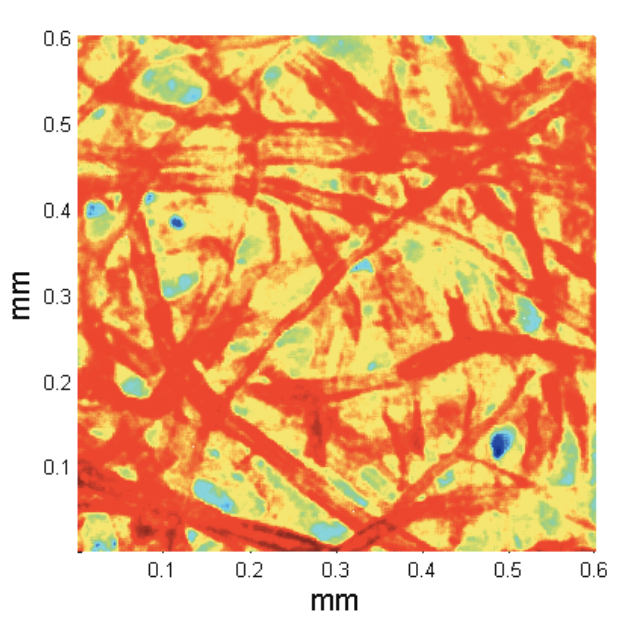}}
    \hfill
    \subfloat[\label{fig:topology}]{%
        \includegraphics[height=3cm,width=0.2\linewidth]{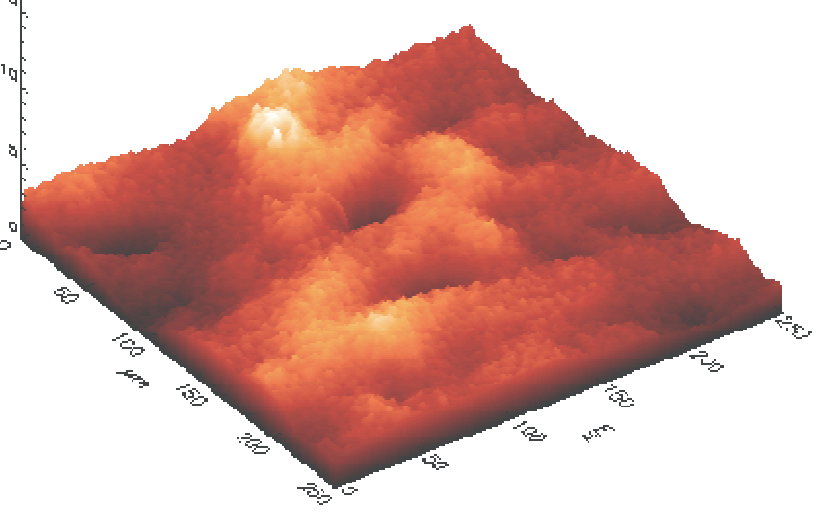}}
    \hfill
    \subfloat[\label{fig:normmap}]{%
        \includegraphics[width=0.2\linewidth]{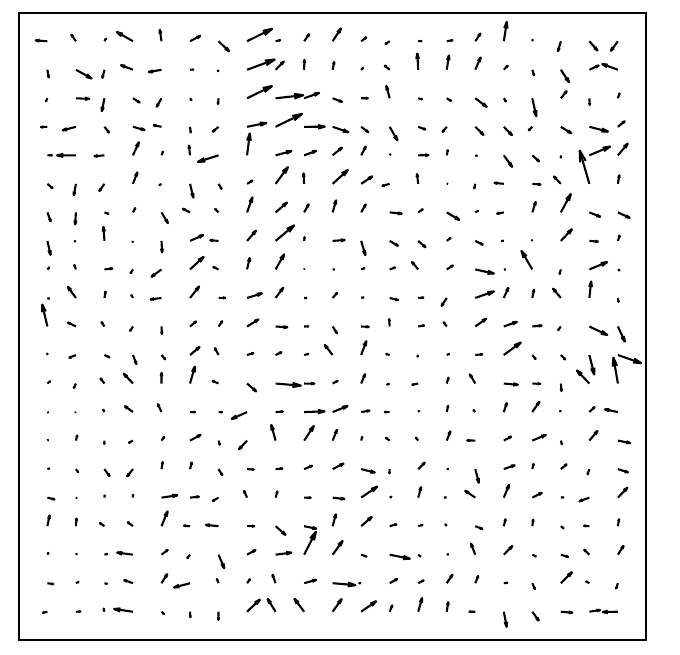}}
  \caption{[Top row] Examples of extrinsic identifiers: (a) Copy detection pattern template (reproduced from \cite{khermaza2021can}), (b) BubbleTag\texttrademark, \ (c) FiberTag\texttrademark, \ (adapted from \cite{bubblefibertag}), and (d) sticker with randomly positioned fibers (adapted from \cite{kindetag}). [Bottom row] Examples of intrinsic identifiers, visualized in (e) A RGB image, (f) a confocal microscope image, and (g) a topography map of paper surface respectively, adapted from \cite{rehberger2007topographical}. (h) Norm map.}
\end{figure*}

In the physical world, the tracking and tracing of a product is frequently done using \textit{extrinsic} identifiers such as bar codes \cite{barcode}, QR codes \cite{qrcode} and radio frequency identification (RFID) tags \cite{angeles2005rfid} that are attached to the product or its packaging. Even though they differ in implementation, all three identifiers can be thought of as an unencrypted container storing data that can be scanned and decoded. Bar/QR codes and RFID tags have severe security shortcomings. Due to their unencrypted design, they are easily duplicable using easily available devices like mobile phones \cite{cloneRFID, counterfeitcdp1}. This implies that counterfeit products can have the same identifier as the authentic product and thus pass through a supply chain undetected. Copy detection patterns (CDP), illustrated in Fig.~\subref*{fig:cdp}, have been added to bar/QR codes to secure against duplication, but these are vulnerable to generative machine learning-based attacks \cite{cdpattack1,slavatifs}. Some active RFID tags support built-in cryptographic protocols but these tags are often expensive and bulky. Finally, these identifiers give an authentication error rate at the order of 0.01 to 0.1 \cite{counterfeitcdp1,counterfeitcdp2} which is too high to be practical for a large-scale supply chain.

Researchers have sought to use special materials with unique visible features. For example, BubbleTag and FiberTag \cite{bubblefibertag} leverage the uncontrollable randomness in the manufacturing process to serve as a signature. BubbleTag, illustrated in Fig.~\subref*{fig:bubbletag}, is a tamper-resistant seal that contains random patterns of bubbles, whereas FiberTag, visualized in Fig.~\subref*{fig:fibertag}, randomly distributes colored fiber on surfaces. These tags are authenticated by using automated computer vision tools or by human inspection. Their authentication performance is adequate for supply chains \cite{wongicip}, but they often require special proprietary infrastructure and are relatively expensive. 

\subsubsection{Intrinsic Identifiers}

The limitations of the extrinsic identifiers motivate a solution that adds minimal overhead to a supply chain, is easy to acquire, and provides authentication performance adequate for supply chains. A viable class of candidates that satisfies all these properties are the \textit{intrinsic} identifiers referred to as optical physically unclonable functions (PUFs). PUFs are an established concept \cite{papputhesis,tuyls2005information}, which have been widely used as primitives for cryptographic protocols such as key distribution \cite{10.1145/1284680.1284683}. In a challenge--response scenario, these functions model light as a challenge signal applied upon physical material to obtain such output responses that are difficult to replicate due to the uncontrollable intrinsic material properties. The idea of an optical PUF was first put forth in \cite{pappu2002physical} where a laser was shined upon an inexpensive, inhomogenous material to obtain random laser speckle patterns. Hammouri et al. \cite{hammouri2009cds} further explore optical PUFs by using laser readers to measure the deviations in the length of pits and lands of compact disk (CD), which is then used as a unique signature.

Complementary to cryptographic studies that focus
on using these intrinsic imperfections as strong sources of randomness \cite{yildiz2020plgakd,huang2017puf,paral2011reliable}, various researchers have explored the possibility of using the inherent randomness as a signature to authenticate objects for anti-counterfeiting purposes. Existing work has largely focused on paper surfaces due to the ease of modeling its interaction with light \cite{clarksonpuf,wongicip,wongtifs,liu2018enhanced}. The extraction of the inherent randomness of the paper surface has been studied by using both physical modeling and visual appearance. \textit{Physical intrinsic features} model the surface texture of the physical material directly. One of the first works to extract the microscopic level texture of a paper surface as a unique signature is by Clarkson et al.~\cite{clarksonpuf} who use commodity scanners to acquire four views of the image at four different orientations. Based on these four images and a fully diffused reflection model, this work estimates a 2D map (referred to as the \textit{norm map}) of the 3D normal vectors. Since flatbed commodity scanners might not always be available, Wong et al. \cite{wongtifs,wongwifs} extend the estimation of the norm map to use the camera of mobile devices. Further work by Liu et al.~\cite{liu2018enhanced} enhanced this method by improving the underlying reflection model. Apart from studies that physically model the surface texture, a significant body of research employs specialized imaging acquisition setups to capture high-resolution visual features of the paper surface \cite{kariakin,beekhof2008secure,smith1999microstructure,guarnera2019new}. However, visual features have limitations as they are brittle to different capturing conditions, may require specialized imaging methods to function well, and often have weaker authentication performance than physical features \cite{clarksonpuf,wongtifs}.

\subsection{The Cyber World}

Although an identifier might be unclonable in the physical world, the verification scheme operates in the cyber world and needs to be secure against adversaries. The cyber world of a supply chain has multiple requirements. First and foremost, there is a need to track a large number of products going through the supply chain. Even though the physical products might have strong, unclonable identifiers associated with them, there needs to be a detailed product tracking backend system attached to them so that the stakeholders always know the location of the product. Second, there needs to be an irrefutable audit trail so that a clear chain of custody can be established. As it calls for multiple stakeholders to collaborate and provide each other the ability to access their sensitive information securely, these needs motivate the use of a secure ledger to record supply chain transactions. 

These ledgers can be implemented through various methods. Since supply chains often track and trace a large volume of products, there is a strong need for throughput-oriented solutions. Thus, relational databases with access control are widely adopted \cite{10.1145/1376616.1376648}, and such databases may perform several tens of thousands of operations per second \cite{ruan2021blockchains}. However, most commercial databases are operated by a centralized authority and are unable to operate in hostile environments. Since supply chains are complex, it is a nontrivial task to decide on the single trusted authority that operates this database. Due to this reason, blockchains have received considerable attention recently \cite{chang2020blockchain,pun2021blockchain,alzahrani2018block} to address security needs and mitigate the drawbacks of traditional databases. They maintain an immutable, append-only record of transactions to prevent data tampering \cite{monrat2019survey}. They do not require central authority and can support access control mechanisms. However, they lack the throughput of traditional databases \cite{croman2016scaling}. Additionally, storing a large amount of data on the chain is common in track-and-trace applications but is considerably difficult. Several key recent works integrate the characteristics of blockchain technology with those of a traditional database. These hybrid systems either add security properties on top of a database \cite{mcconaghy2016bigchaindb,nathan2019blockchain}, or improve the throughput of a blockchain by adding database capabilities \cite{el2019blockchaindb,10.1145/3318464.3380594,allen2019veritas} based on the desired application.

%% file: Sections/sec3_physicalpaperPUF.tex
\section{Protecting the Physical World Using Intrinsic Paper PUF Identifiers\label{Sec3}}

The limitations of extrinsic PUFs discussed in the last section motivate us to consider intrinsic PUFs as a scalable alternative. Additionally, the economical nature, the simplicity of physically modeling interactions with light, and the abundance of prior studies inspire us to consider paper-based anti-counterfeiting solutions. Although micro-level intrinsic features extracted from the paper surface are often assumed to be unclonable \cite{toreini2017texture,clarksonpuf,wongicip,wongtifs,liu2018enhanced}, we are not aware of any past studies that explicitly justify why this is true. In this section, we justify this assumption by examining the paper-making process and identifying the key step that introduces randomness into the microscopic texture of a paper surface. We further justify the use of micro-level paper features as an optical PUF by analyzing their ability to meet cryptographic PUF properties.

\subsection{Origin of Paper's Unclonable Microstructures}

Paper-making is an elaborate process that combines different intricate chemical and mechanical steps \cite{sjostrom1998analytical,philpott2001supply}. We abstract the complex paper-making process into six steps. (1) The first step is to source the raw materials, for instance, raw wood from lumber mills or recycled paper from material recovery facilities. (2) Once the raw materials are sourced, they are broken into smaller, more usable materials. (3) Then at pulp mills, the fibers are extracted from these materials through a chemical and mechanical process called pulping. (4) These fibers are washed and refined. This enhances the properties of the individual fibers such as their tensile strength and smoothness. (5) Next, these fibers are shaped into a sheet of paper through the process of forming, wherein the pulp slurry is dropped onto a wire mesh and left to form. During forming, the paper fibers are left to intertwine with each other and they entangle non-deterministically. (6) After a sheet of paper is formed, the paper is dried and further processed through bleaching and other methods.

Uncontrollable differences arise between different paper sheets at every stage of the pipeline. Different sheets may be crafted from different batches of raw materials thereby adding randomness. Additionally, different pulping and refining techniques might be used for different sheets. However, one can assume that these techniques are similar at the same paper mill, thus not adding much randomness to the microstructures on average. The largest amount of randomness is introduced during the process of forming. In forming, the fibers sit on a mesh, physically interact with each other, and intertwist, a process that does not admit a compact mathematical representation. Hence, it is due to forming that the PUF properties of paper surfaces arise.

\subsection{Desired Properties for Paper PUFs \label{pufproperties}}

To understand the suitability of paper microstructures as an optical PUF, we should check if they satisfy standard PUF metrics such as those given in \cite{maiti2013systematic,katzenbeisser2012pufs}. While these metrics usually evaluate the randomness and reliability of PUF responses for use as a one-way function, they are also crucial for designing a robust and secure authentication system. This is because the suite of metrics informs a system designer about the feature variations, which shine a light on the feasibility of using a feature for authentication. In this subsection, we provide a few key PUF evaluation metrics to examine the feasibility of the norm map as a paper PUF-based authentication feature.

We adopt the following notation: (1) An evaluated paper-PUF is associated with the $k$th product out of $K$ distinct products, (2) Each PUF is evaluated $T$ times in different conditions, (3) Each PUF response ${\bf r}^{k, t} \in \{0, 1\}^L$ has $L$ bits, and (4)~$\mathrm{HD}({\bf x},{\bf y})$ refers to the Hamming distance between vectors ${\bf x}$ and ${\bf y}$. We consider four PUF-based evaluation metrics:

\begin{enumerate}

    \item \textit{Robustness}: This measures the stability of the output under different evaluation conditions such as illumination changes: Ideally, the responses should be the same regardless of variations in the acquisition, and thus, the ideal value of robustness is 1.

    \begin{equation}
        \mathrm{Robustness}(k) = 1-\!\frac{1}{T} \sum_{t=1}^{T} \frac{\mathrm{HD}({\bf r}^{k, \text{ideal}},{\bf r}^{k, t})}{L}.
    \end{equation}

    \item \textit{Uniqueness}: This measures the discriminative power of the PUF.
    \begin{equation}
        \mathrm{Uniqueness}(N) \! = \!\frac{1}{\binom{K}{2}} \! \sum_{k=1}^{K-1} \! \sum_{k'=k+1}^{K} \! \frac{\mathrm{HD}({\bf r}^{k, t},{\bf r}^{k', t})}{L}.
    \end{equation}

    It quantifies the ability to distinguish a response from any other response. An ideal value of 0.5 implies perfect discriminability between any given pair of PUFs. 

    \item \textit{Uniformity}: This measures uniform randomness of the 0s and 1s in a PUF response: 
    
    \begin{equation}
        \mathrm{Uniformity}(k,t) = \frac{1}{L} \sum_{\ell=1}^{L}r^{k, t}_\ell.
    \end{equation}
    
    It quantifies how distinguishable the response is from a truly uniformly random sequence. Uniformity should ideally be 0.5 for the 0s and 1s to be equiprobable. 
    
    \item \textit{Ease of Evaluation}: This qualitative metric considers the ease of generating a PUF response. 

\end{enumerate}

Even though no work thoroughly tests all of these metrics on deployable anti-counterfeiting systems, paper-PUF has shown promise to satisfy these properties. First, if we consider norm map as the feature of choice, \textit{robustness} is satisfied because such estimation methods as in \cite{wongtifs,liu2018enhanced} compensate for various variations including illumination conditions in the preprocessing. Second, norm maps exhibit \textit{uniqueness} since the EER shown in \cite{wongtifs,liu2018enhanced} is of the order $10^{-100}$ to $10^{-10}$. Third, norm maps can be considered to be globally \textit{uniform} even if they are not locally. Nearby locations (pixels) in a norm map are weakly correlated due to the diffusion nature of light and the intertwisting of nearby fibers. However, the regions that are moderately further away (5--10 pixels) can be assumed to be uncorrelated since the texture information overshadows any small correlation arising due to other factors. Fourth, the work in \cite{wongtifs,clarksonpuf,liu2018enhanced,flatbed} demonstrate \textit{ease of evaluation} by capturing regions of interest of paper using equipment such as flatbed scanners or mobile cameras to estimate the norm map. These properties make norm map-based anti-counterfeiting systems suitable for deployment.

\subsection{Physical Attacks\label{physicalattack}}

Even though the paper-PUF response is physically unclonable, several challenges still exist. If an adversary's goal is to make authentication impossible rather than to pass a counterfeit product off as genuine, a variety of denial-of-service (DoS) attacks are possible. An adversary can try to destroy the texture of the paper surface by crumpling, wetting, printing or scribbling on top of the paper patch used to authenticate a product. Toreini et al.~\cite{toreini2017texture} perform an ablation on image features under a variety of such threats and find that the robustness of attacked paper patches drops but the uniqueness in terms of discriminative power remains stable. Clarkson et al.~\cite{clarksonpuf} evaluate the feasibility of using norm maps under a similar set of attacks and also reported no significant loss of \textit{uniqueness}. However, both studies did not consider heavily destroyed patches and do not consider the more challenging cases such as tearing and scratching. Similar to a tamper-proof seal that shows clear signs of manipulation, these physical attacks are visually obvious and can be used to flag potential counterfeit products. It is however important to understand the differences between naturally degraded surfaces and attacked surfaces, a critical aspect that has not been explored in the current literature. Overall, considering both the strengths and the limitations, paper-PUF-based authentication systems are promising for real-world applications.

%% file: Sections/sec4_cyberframework.tex
\section{A General Framework for Paper-based Authentication Systems \label{Sec4}}

\begin{figure*}[!t]
    \centering
    \includegraphics[width=\linewidth]{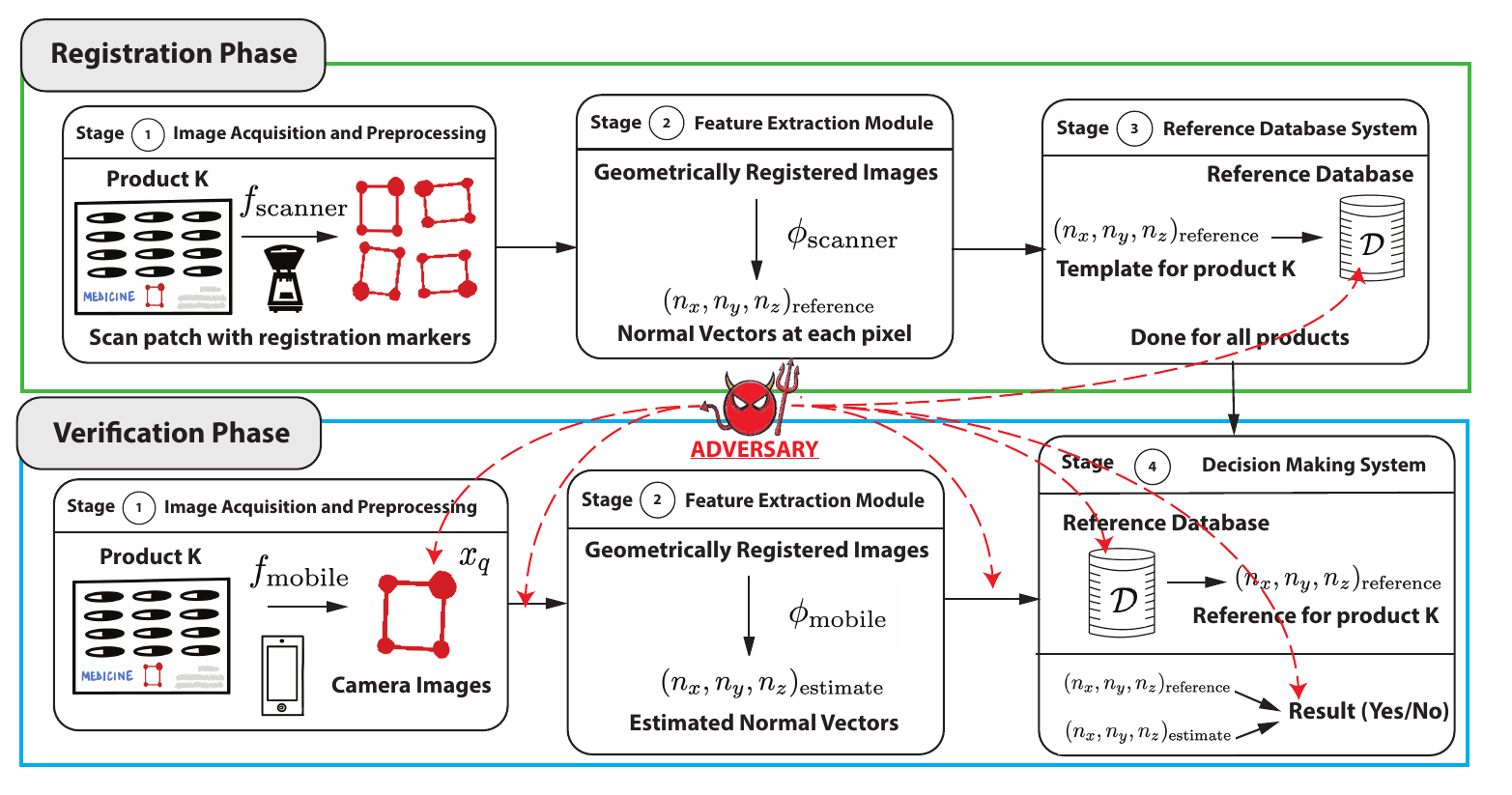}
    \caption{The norm map based anti-counterfeiting system framework that divides $\mathcal{A} = (f,\phi,\mathcal{D},\delta) $ into different operational stages 1--4. We highlight the potential attack location of significant threats using red dashed arrows.}
    \label{fig:framework}
\end{figure*}

To establish a trusted cyber--physical link to combat counterfeits, it is essential to understand the security of the cyber and physical worlds together. In this section, we consolidate the interactions of the physical and cyber worlds of a supply chain into an authentication framework.

\smallskip
\noindent
\textbf{Notation/Terminology:} In this paper, we denote the paper-PUF authentication system as $\mathcal{A} =(f,\phi,\mathcal{D},\delta)$ where $f$ refers to the preprocessing function that converts a query $x_{\text{q}}$ into a form appropriate for feature extraction, $\phi$ converts the preprocessed query $f(x_{\text{q}})$ into an authentication feature vector, $\mathcal{D}$ contains the reference feature vectors, and $\delta$ is the decision rule that compares $\phi(f(x_{\text{q}}))$ to the references in $\mathcal{D}$.

\subsection{Norm Map Based Authentication}

We now examine the different parts of the authentication system $\mathcal{A}$ in detail. The authentication pipeline can broadly be divided into four different stages as shown in Fig. \ref{fig:framework}. These stages form the core operations of the authentication system and are processes where a supply chain stakeholder interacts with the system. These stages and their interactions also form the attack surface of the authentication system where an adversary will attack and try to take control. Similar to most biometric systems, a paper-PUF-based authentication system operates in two phases. The first phase is referred to as the \textit{registration} phase. In this phase, the manufacturer in the supply chain extracts template norm maps from images of the surface of a product (or its paper packaging) and stores them in the reference database $\mathcal{D}$. In the second phase, known as the \textit{verification} phase, the other stakeholders send queries ($x_{\text{q}}$) to $\mathcal{A}$ to authenticate the various products they possess. As an example, we now explain the different operational stages in the context of a norm-map-based anti-counterfeiting system. 

\smallskip
\noindent
\textbf{Stage 1: Image Collection and Preprocessing using $f$.} This stage collects and preprocesses multiple images of a paper patch, which are then to be fed into norm map estimation algorithms. For scanned patches, the estimation of normal vectors requires the acquisition of four images scanned at 0\textdegree, 90\textdegree, 180\textdegree\ and 270\textdegree\ as outlined in \cite{clarksonpuf}. If the acquisition is done using mobile devices as in \cite{wongtifs}, four or more images are required to solve the under-determined system of linear equations. The preprocessing for the norm map includes geometric registration, which is necessary since any slight misalignment causes a significant drop in performance.

\noindent
\textbf{Stage 2: Feature Extraction via $\phi$.} The feature extraction module implements an estimator based on a fully diffused reflection model to extract norm maps from the preprocessed images. When using scanners \cite{clarksonpuf}, the difference between images scanned at 0\textdegree \ and 180\textdegree\ is proportional to the $y$ component of the normal vector $n_{y}$ and the difference between the images scanned at 90\textdegree \ and 270\textdegree\ is proportional to the $x$ component of the normal vector $n_{x}$. When using a camera, $n_{x}$ and $n_{y}$ are estimated by a least-squares solver \cite{wongtifs}.

\noindent
\textbf{Stage 3: Storing Reference Data in $\mathcal{D}$.} The storage solution~$\mathcal{D}$ stores the template norm maps acquired during the registration phase. These references will be retrieved during the verification phase by the decision-making system to assist in the computation of an authentication decision. Since scanners offer a better resolution of the microscopic texture of the surfaces, the templates can be registered into the database using a scanner in the registration phase. The reference norm map is then stored as a (ID, Template) pair so that matching can be done based on ID. Alternatively, the reference templates can also be stored without any corresponding ID, and in this case, the verification process will perform a similarity search over $\mathcal{D}$.

\noindent
\textbf{Stage 4: Decision-Making using $\delta$.} The decision-making stage is responsible for generating a yes/no result signifying the success or failure of the authentication. For the $k$th product, this stage computes $\phi(f(x_{\text{q}}))$. It is then compared against $\phi(f(x_k))$ if the template is stored with the product ID, or otherwise against $\{\phi(f(x_{i}))\}_{i=1}^N$ where $N$ is the total number of templates in the reference database $\mathcal{D}$. The comparison may be done using distance metrics such as the $\ell_{2}$ distance or the Pearson correlation ($\rho$) e.g, $\delta(x_{\text{q}},x_{k},\epsilon)=\mathds{1}(\|\phi(f(x_{\text{q}})) - \phi(f(x_k))\|_{2} \leq \epsilon) $ for a chosen similarity threshold $\epsilon$.

%% file: Sections/sec5_taxonomy.tex
\section{Taxonomy of Existing Methods\label{Sec5}}


\begin{table*}[!t]
\caption{Taxonomy of different Anti-Counterfeiting Systems based on the framework in Fig \ref{fig:framework}. \label{taxonomy}}
\resizebox{\linewidth}{!}{%
\tabcolsep=0.11cm
\setlength\extrarowheight{0.6mm}
\begin{tabular}{|l|l|l|l|l|l|l|p{1cm}|l|}
\hline
\multirow{3}{*}{\textbf{Ref.}} & \multirow{3}{*}{\textbf{Year}} & \multicolumn{6}{c|}{\textbf{Implementation}} & \multirow{3}{*}{\parbox{1cm}{\textbf{Is it \\ Deploy-\\able?}}} \\ \cline{3-8}
 & & \multicolumn{3}{c|}{\textbf{Physical World}} & \multicolumn{3}{c|}{\textbf{Cyber World}} &  \\ \cline{3-8}
 & & Signal Acquisition ($f$) & Feature Details ($\phi$) & Materials Considered & Storage ($\mathcal{D}$) & Verification Logic ($\delta$) & Analyze \newline Security? & \\
\hline \hline

\multicolumn{9}{c}{\textit{Methods Using Specialized Signal Acquisition Setups}} \\ \hline \hline 

\cite{pappu2002physical} & 2002 & Laser \& CCD Camera & Gabor filtered Image & Epoxy Tokens &  Database & Hamming Distance & \checkmark & \text{\sffamily X}\\
\cite{buchananlitreview} & 2005 & Laser \& Photodetectors  & Intensity fluctuations & Paper &  ? & Digital Cross-Correlation & \text{\sffamily X} & \text{\sffamily X}\\
\cite{cowburn2008laser} & 2008 & Laser \& Photodetectors & Laser Speckle Scans & Paper &  ? & Digital Cross-Correlation & \text{\sffamily X} & \text{\sffamily X} \\
\cite{sharma2011paperspeckle} & 2011 & USB Microscope w/ LED & Gabor filtered speckle & Paper &  Offline & Fractional Hamming Distance & \checkmark & \checkmark \\ 
\cite{chenandzeng} & 2020 & Bimodal Camera & Gabor filtered Image & Paper &  Database & Hamming Distance & \text{\sffamily X} & \text{\sffamily X}\\ 
\cite{kumar2022unclonable} & 2022 & Lasers \& CCD/Mobile & Image of MLA label  & Any &  ? & Pixel Intensity Comparison & \text{\sffamily X} & \text{\sffamily X} \\ \hline \hline

\multicolumn{9}{c}{\textit{Methods Using Commonly Available Visual Acquisition Setups and Visual Features}} \\ \hline \hline 

\cite{kariakin} & 1997 & Scanner & Diff lighting angle Images & Paper &  On Banknote & Hamming Distance  & \checkmark & \text{\sffamily X} \\
\cite{smith1999microstructure} & 1999 & Video Camera & Preprocessed Image & Paper &  ? & Cross-Correlation & \text{\sffamily X} & \checkmark \\
\cite{bubblefibertag} & 2005 & Camera & Image of designed tag & Any &  Database & ? & \text{\sffamily X} & \text{\sffamily X}\\
\cite{beekhof2008secure} & 2008 & Mobile Camera, Macro lens & Preprocessed Images & Paper &  Database & Min. Reference Distance & \checkmark & \checkmark \\
\cite{metalpuf} & 2012 & Digital Imaging Device & Preprocessed Image & Paper, Metal, Resins &  Database & Cross-Correlation & \text{\sffamily X} & \checkmark \\
\cite{toreini2017texture} & 2017 & Digital Imaging Device & Gabor filtered Image & Paper &  Database & Fractional Hamming Distance & \checkmark & \checkmark \\
\cite{guarnera2019new} & 2019 & Projector, RGB Camera & LBP Histogram & Paper &  Database & Bhattacharyya Distance & \text{\sffamily X} & \checkmark\\ \hline \hline

\multicolumn{9}{c}{\textit{Methods Adding Extrinsic Identifiers to Paper  }} \\ \hline \hline 

\cite{boschcounterfeit} & 2018 & Digital Camera & Barcode from Image & Any &  Database & Serial Number Matching & \checkmark & \checkmark\\
\cite{chen2019copy} & 2019 & Mobile Camera & DFT,LBP of barcode & Paper &  SVM Model & SVM output & \text{\sffamily X} & \text{\sffamily X}\\ 
\cite{yan2020iot} & 2020 & Mobile Camera & AKAZE, printed microdots & Paper &  Database & H.D + Salient-Point Matching & \checkmark & \checkmark\\ \hline \hline

\multicolumn{9}{c}{\textit{Methods Using Intrinsic Norm Map as a Feature}} \\ \hline \hline 

\cite{clarksonpuf} & 2009 & Flatbed Scanner &  Norm map (N.M.) & Paper &  Database & Hamming Distance (H.D)  & \checkmark  &  \checkmark \\
\cite{wongtifs} & 2017 & Mobile Camera, Scanner & N.M. & Paper & Database & Cross-Correlation & \text{\sffamily X} & \checkmark \\
\cite{liu2018enhanced} & 2018 & Mobile Camera, Scanner & 3D Height Map \& N.M. & Paper &  Database & Cross-Correlation & \text{\sffamily X} & \checkmark \\
\cite{flatbed} & 2021 & Flatbed Scanner & N.M. & Paper &  Database, Offline & Cross-Correlation & \checkmark & \checkmark\\ \hline \hline

\multicolumn{9}{c}{\textit{Methods Using Blockchain-based Storage Solutions}} \\ \hline \hline 

\cite{toyoda2017novel} & 2017 & RFID Scanner & RFID Data & Any &  Blockchain & Electronic Product Code & \checkmark &  \text{\sffamily X} \\
\cite{alzahrani2018block} & 2018 & NFC Tag Reader & NFC Tag ID & Any &  Blockchain & NFC Data &  \checkmark &  \text{\sffamily X} \\
\cite{kumar2019traceability} & 2019 & QR Code Scanner & QR Code data & Any &  Blockchain & QR Decoded Data & \checkmark & \text{\sffamily X}\\
\cite{aniello2019towards} & 2019 & Ring Oscillator PUF Tags & PUF Response & Any &  Blockchain & Perfect Response Match & \checkmark & \checkmark\\ \hline \hline

\multicolumn{9}{c}{\textit{Methods That Investigate Other Materials Than Paper}} \\ \hline \hline 

\cite{FIBAR} & 2014 & Mobile Camera, Macrolens & Preprocessed Image & Metal &  Database & RANSAC Inliers + ORB & \text{\sffamily X} & \checkmark\\
\cite{ishiyama2016midot} & 2016 & Modified FIBAR \cite{FIBAR} & Preprocessed Image & Metal, Plastic &  Database & RANSAC Inliers + ORB & \text{\sffamily X} & \checkmark\\
\cite{sharma2017fake} & 2017 & Proprietary Microscope & Dense SIFT & Leather, Fabric, Plastic &  CNN Model & CNN Output & \text{\sffamily X} & \text{\sffamily X} \\
\cite{ishiyama2018fast} & 2018 & Microscopic Imaging & FMT of Image & Metal &  Database & DFT Correlation & \text{\sffamily X} & \checkmark\\
\cite{wigger2018label} & 2018 & Camera, Special Flash & Preprocessed Image & PCB Samples &  Database & openCV Template Matching & \text{\sffamily X} & \checkmark \\ 
\cite{wigger2020robust} & 2020 & Video Measuring System & DCT Hashed Image & MID Plastic &  Database & Hamming Distance & \checkmark & \text{\sffamily X}\\
\cite{wang2021anti} & 2021 & Negative Film Scanner & Gabor filtered Image & Polymer Banknote &  Database, Offline  & Fractional Hamming Distance & \checkmark & \text{\sffamily X} \\ \hline
\end{tabular}
}
\\
\end{table*}

Using our anti-counterfeiting framework, we develop a comprehensive taxonomy of 31 different anti-counterfeiting methods proposed over the last three decades and summarize them in Table \ref{taxonomy}. The taxonomy is developed based on the physical and cyber world solutions that are proposed in the respective works. In the physical world, the works are categorized based on the used signal, its corresponding acquisition equipment, the type of authentication feature considered, and the different physical materials that were tested. In the cyber world, the type of storage solution $\mathcal{D}$, the matching logic $\delta$, and if the works do any security analysis is considered. Additionally, based on the physical and cyber world solutions of the different systems, we briefly evaluate their deployability. This evaluation is based on different desirable properties that we outlined prior to this section such as the non-proprietary nature of design, unclonability, and the use of standard, widely available equipment for instance, mobile cameras or commercial scanners.

Using our taxonomy, we uncover several key findings. First, there is a significant amount of work that considers intrinsic features of paper but a very limited amount of work expands ideas to other materials that may be what products are made from such as fabric, metals, or plastics. Second, it is apparent that there is a dichotomization in the type of analysis of the different systems. Prior works either concentrate on designing a robust, unclonable physical world and do not analyze the cyber world extensively, or they create secure cyber worlds but use clonable extrinsic physical identifiers. Third, most works consider \textit{visual} features rather than a more general class of \textit{physical} features. These features are deployable due to their use of ubiquitous sensing mechanisms such as mobile cameras and simple verification logic. However, their sensing mechanisms are sensitive to variations in environmental conditions such as illumination changes. Finally, in the paper-based anti-counterfeiting systems, norm-map-based systems stand out. They use accessible and ubiquitous means of image acquisition, use physical intrinsic features that are unclonable, do not add a significant cost overhead, and do a preliminary security analysis through threat modeling. These characteristics coupled with very low authentication error rates and PUF-like properties make them a potent solution for anti-counterfeiting applications.

%% file: Sections/sec6_sysdesign.tex
\section{Attacks on paper-PUF-based Authentication\label{Sec6}}

\subsection{Desired Properties of Secure Systems}

Before service providers or stakeholders of the supply chain design a paper-PUF-based authentication system, it is essential to understand the key desirable security properties. We outline them as follows:

\begin{enumerate}
\item \textbf{Confidentiality}: Confidentiality deals with the covertness of a feature, for example: protecting the reference feature against leakage.
\item \textbf{Integrity}: This refers to the tamper resistance of a feature so that authentication results can be trusted. For example: securing reference features in the database.
\item \textbf{Revocability}: Since intrinsic identifiers are the ``biometrics" of the products, it is critical to secure them by using cancelable and re-enrollable hash methods.
\item \textbf{Replay Resistance}: The verification system should be protected against adversaries who may try to attack the channels between the different operational stages.
\item \textbf{Availability}: Since even a small system outage can cause large losses for supply chains, the availability of the system is important. Being distributed and incorporating redundancy is crucial to mitigate DoS attacks. 
\end{enumerate}

Additionally, different stakeholders in a supply chain may not have a priori information about other stakeholders' intents. In such a scenario without established trust, it is essential to understand how stakeholders can collaborate to ensure the proper functioning of the supply chain. We discuss these practical considerations in Section~\ref{Sec7}

\subsection{Vulnerabilities of Consolidated Framework\label{attackcommon}}

In this section, we systematically explore the various vulnerabilities of the system $\mathcal{A} = (f,\phi,\mathcal{D},\delta)$ by using the framework illustrated in Fig.~\ref{fig:framework}. In the ensuing subsections, we state the key threats to each operational stage in this framework, the security properties that each threat targets, and provide a set of representative mitigation strategies. Unless otherwise stated, we study the \textit{verification phase}. We summarize the system properties targeted by each stage in Table~\ref{fig:proptable}. 

\begin{table}[t!]
    \caption{\label{fig:proptable} Properties targeted ($\bullet$) and not targeted (-) in the stage-wise threat modeling of the consolidated framework}
    \resizebox{\linewidth}{!}{%
    \centering
    \renewcommand{\arraystretch}{1.2}
\begin{tabular}{P{1.5cm}*{5}{c}}
                                        \toprule
    &   \multicolumn{5}{c}{\thead{System Properties}}
                                    \\  \cmidrule{2-6}
\thead{Attacked \\ Stage}
    &   {\thead{Confident-\\iality}} 
        &   {\thead{Integrity}} 
            &   {\thead{Revocab-\\ility}} 
                &   {\thead{Replay \\ Resistance}}
                    &   {\thead{Availability}}
                                    \\ \midrule
\centering 1
    & - & - & - & - & \Large \textbullet            \\
\centering 2
    & - & \Large \textbullet  & - & - & -            \\
\centering 3
    & \Large \textbullet & \Large \textbullet & \Large \textbullet & - & \Large \textbullet            \\
\centering 4
     & \Large \textbullet & - & - & - & -            \\
\centering Interface
    & \Large \textbullet & \Large \textbullet & - & \Large \textbullet & -           
                                    \\  \bottomrule
    \end{tabular}}
\end{table}

\subsubsection{Attacks on Stage 1: Image Acquisition}
\hfill \smallskip \\
This stage uses different sensors such as flatbed scanners and mobile cameras to capture images of the paper surface and then preprocesses them for feature extraction. We start by examining the most effective threats that can plague this stage. 

\smallskip
\noindent \textbf{Properties targeted}: \textit{Availability; PUF Ease of Evaluation}

\smallskip
\noindent \textbf{DoS attacks}: Multiple works \cite{kariakin,wongicip,wongtifs,wongwifs,liu2018enhanced,flatbed} use image registration patches (as illustrated in Fig.~\ref{fig:framework}) to align various snapshots obtained under different capturing conditions. Exploiting this insight, an adversary can physically tamper with the registration markers to worsen the performance of the authentication system. Since preprocessing images is a major step for subsequent stages, this would imply that the authentication feature can be rendered unusable. More challenging scenarios mentioned in Section \ref{physicalattack} such as tearing and scratching of paper surface are also easy to carry out and can make authentication impossible. Alternatively, an adversary may also physically sabotage the imaging sensor such as a mobile camera.

\smallskip
\noindent \textbf{Sensor Spoofing}: Rather than denying service to supply chain stakeholders, an adversary can potentially aim to provide an authentication record for a counterfeit product by incorrectly authenticating it as legitimate. This resembles the spirit of sensor spoofing, which has been well-studied in the field of biometrics \cite{schuckers2002spoofing,hadid2015biometrics,menotti2015deep}. Spoofing is accomplished by generating fake input samples that generate artificially high decision scores, hence passing authentication. For anti-counterfeiting applications, spoofing has not been comprehensively studied. Extrinsic identifiers such as 1D/2D codes or RFID tags are often convenient to clone. This means that an adversary can replace these identifiers with their cloned counterparts to authenticate a counterfeit product. Alternatively, intrinsic identifiers such as norm maps use images of paper surfaces to extract authentication features. Several image-domain spoofing attacks have been studied against face-recognition biometric systems and are feasible for paper-based authentication systems \cite{kumar2017comparative,wu2020making}. If an adversary is able to obtain images of the intrinsic identifier that is associated with a genuine product, they can simply feed these images into the sensor and circumvent authentication. 

\smallskip
\noindent \textbf{Mitigation Strategies}: DoS attacks and spoofing attacks are categorized as \textit{direct} attacks \cite{hadid2015biometrics} and are hard to defend against. Since the DoS attacks visibly change the appearance of the identifier, quality control checkpoints can isolate them with ease. Spoofing attacks are harder to defend against because they are carried out at the sensor level. Face biometric spoofing attacks are often defended against by using the intricacies of the acquisition process itself. Liveness detection are techniques that detect signs of life and often form the backbone of anti-spoofing literature for biometrics \cite{pan2008liveness}. However, the inanimate nature of paper renders these techniques inappropriate for our purposes. Image forensics strategies such as evaluating the metadata of the input images or requiring multiple snapshots of the identifiers \cite{fridrich2009digital} are promising solutions for dealing with these attacks in a supply chain. Leveraging combinations of different identifiers, akin to multimodal matching, can also mitigate spoofing attacks \cite{wild2016robust}.

\subsubsection{Attacks on Stage 2: Feature Extraction}
\hfill \smallskip \\
The second stage of the pipeline is responsible for extracting the authentication features from the preprocessed images outputted by stage 1. The attacks in the first stage have a direct and strong influence on the feature extraction process. DoS attacks can render any output produced by this stage meaningless due to improper inputs to the algorithms of this stage. Even if the system designer is able to protect the prior stage, an adversary can attack this stage in isolation. 

\smallskip
\noindent \textbf{Properties targeted}: \textit{Integrity; PUF Unclonability}

\smallskip
\noindent \textbf{Surrogate Models}: Modeling surrogates has received considerable attention in the field of machine learning. Given a target machine learning system and its paired input-output data, this attack aims to learn the underlying mathematical representation that maps the input space of a model to its output space. Such attacks have been well explored in the field of adversarial machine learning \cite{ilyas2018black,guo2019simple,narodytska2017simple,bhambri2019survey}, where adversarial attacks crafted on such learned representations often transfer to the target model \cite{papernot2016transferability}, thereby rendering the target model vulnerable. 

Machine-learning-based modeling attacks have also been explored in the context of arbiter PUFs \cite{pufattack2,ruhrmair2013puf}. These studies find that PUFs are susceptible to such attacks, as simple models such as logistic regression are enough to learn a mathematical representation that emulates their delay behavior. However, this representation might not be the true underlying representation, but rather an overfitted mapping. Although existing literature does not explore modeling attacks against paper PUFs, prior work on arbiter PUFs and the ability of machine learning algorithms to learn complicated, seemingly intractable functions makes such attacks potentially feasible against intrinsic identifiers such as norm maps. If an adversary is able to learn a general well-fitted mathematical representation, then the unclonability of these identifiers is lost and the system is made vulnerable.  

\smallskip
\noindent \textbf{Synthetic Feature Generation}:
Another machine-learning-based attack that can compromise the security of this stage is synthetic data generation. The past decade has seen the rise of generative models that learn the underlying latent space distribution of images. These distributions are then sampled to generate new semantically and visually realistic images. Since many methods leverage visual or image-like physical features for authentication, generative models such as diffusion models \cite{ho2020denoising} may conceivably be used to generate and feed fake features that pass the authentication process.

\smallskip
\noindent \textbf{Mitigation Strategies}: Sophisticated machine learning methods such as neural networks often require a large amount of paired training data. This can be a deterrent for synthetic feature generation. However, various modeling attacks have been carried out with simpler methods such as logistic regression models which do not require much data to train well. A strategy to prevent both modeling and synthetic generation can be to add an \textit{invisible} feature forensics module. Different generative models have been shown to introduce characteristic traces into its generated images such as spatial domain artifacts \cite{corvi2023detection}. This invisible module can be specifically modeled to detect such artifacts and check if the extracted feature is synthetic or not. Additionally, to stop an adversary from collecting a large amount of paired data by repeatedly querying the feature extraction module and observing the output, a lockout mechanism limiting the number of queries can be deployed.   

\subsubsection{Attacks on Stage 3: Reference Database}  
\hfill \smallskip \\
The third stage of the pipeline is the reference database that contains the template features the extracted features are compared to. Securing this stage is essential since it is the ground truth data used for decision-making.

\smallskip
\noindent \textbf{Properties targeted}: \textit{Confidentiality; Integrity; Revocability; Availability; PUF Uniformity}

\smallskip
\noindent \textbf{DoS attacks}: An adversary can simply opt for a DoS attack such as resource overloading \cite{adler2008biometric}.

\smallskip
\noindent \textbf{Template Leakage}: This is an \textit{extraction} attack in which the adversaries gain illegitimate access to the reference database and extract the template feature corresponding to the genuine product. This template feature can then be replayed to the decision-making system, easily bypassing authentication.

\smallskip
\noindent \textbf{Malicious Registration}: If the adversary gets unauthorized access to the reference database, they can generate a new feature corresponding to the counterfeit item, associate it with its corresponding product ID, write it to the reference database, and allow the system to operate as is. Similar actions can be done in the registration phase, as a form of \textit{backdoor} attack.

\smallskip
\noindent \textbf{Inversion Attack}: If the adversary is not able to feed the template directly to the decision-making system (i.e. integrity of channels is preserved), an adversary needs to create input images that generate the leaked template. This is an inversion problem. If the adversary is able to invert the leaked template into the input images, then they can pass these input images into the authentication system and get authenticated. PUF responses are usually assumed to be non-invertible and unclonable due to the inherent randomness in their generation process. However, if learning a compact representation is feasible via machine learning methods or otherwise, then this non-invertibility might not be guaranteed. Since we consider norm maps to be practically deployable, we analyze their invertibility next.

\begin{figure} 
    \centering
        \subfloat[\label{fig:inversion}]{
        \includegraphics[width=0.8\linewidth]{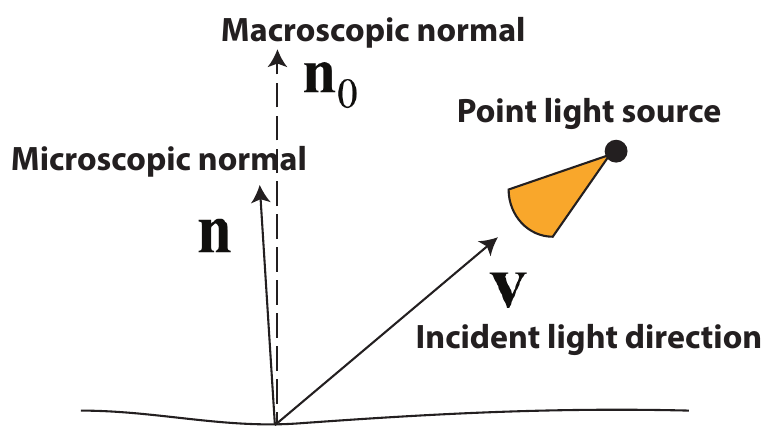}}      
    \caption{(a) A model of the microscopic view of paper used for inversion, adapted from \cite{wongtifs}.}
\end{figure}

\smallskip
\noindent \textbf{Invertibility of Norm Maps\label{NMinvert}}: At its core, a norm map estimation algorithm has two components: (1) It first approximates an environment model and then, (2) solves a deterministic algorithm that returns the normal vectors. For a camera-based norm map estimation, as proposed in \cite{wongtifs}, a diffused reflection model is used where the pixel intensity $l_\text{r}$ at a pixel $\mathbf{p}$ can be written as: 

\begin{equation}
 \\
 l_\text{r} = \lambda \cdot l_0 \cdot {\mathbf{n}^{\top}} \mathbf{v} \big/ \| \mathbf{v} \|^{3}_{2},
\label{eq: diffuse-reflection}
\end{equation}

\noindent where $\lambda$ is a constant albedo (characterizing the light reflective capability), $\mathbf{n}$ is the three-dimensional normal vector $(n_x,n_y,n_z)$ representing the microscopic normal direction, $\mathbf{v}/\|\mathbf{v}\|$ is the direction of incident light from the light source to $\mathbf{p}$ and $l_0/\|\mathbf{v}\|^{2}$ is the incident light intensity at $\mathbf{p}$ using the inverse square law. A visualization of the microscopic view at a spot on the paper surface is illustrated in Fig.~\ref{fig:inversion}.

If an adversary leaks a template from $\mathcal{D}$, they have perfect knowledge of the normal vectors $\mathbf{n}$. To solve for $l_\text{r}$, the adversary would need to estimate the environment model i.e. $l_0$ and $\mathbf{v}$. This is a difficult but tractable problem. For example, a scientist with domain expertise may be able to infer good approximations. Thus, it is possible that an adversary can subsequently solve the deterministic algorithm and recover the raw intensity $l_\text{r}$ at each pixel $\mathbf{p}$ of the norm map to create a synthetic image. Therefore, norm maps are vulnerable to inversion. 

\smallskip
\noindent \textbf{Mitigation Strategies}: Securing against template tampering, template leakage, and malicious registration requires protecting the reference database against intrusion and tampering. This can be achieved in multiple ways. First, the service provider responsible for setting up the authentication infrastructure should follow best practices for access control. However, if the adversary is an insider, this solution is not sufficient. The system designer should potentially choose not to store the reference as an (ID, Template) tuple but rather employ an efficient approximate nearest neighbor data structure with ID being hidden information. This would ensure that the adversary would not be able to do a simple ID lookup for template extraction. The best security practice to overcome the above-mentioned attacks is to secure the database by creating an immutable audit trail that records the operations of the database and is visible to all stakeholders who are part of the supply chain. This can be achieved by using blockchains or blockchain-database hybrids instead of traditional reference databases \cite{mcconaghy2016bigchaindb,nathan2019blockchain,schuhknecht2019chainifydb,el2019blockchaindb,10.1145/3318464.3380594,allen2019veritas}. 

By leveraging key concepts from biometrics and other security practices such as salting \cite{teoh2006random}, revocable non-invertible hashing \cite{teoh20072}, and biometric cryptosystems \cite{davida1998enabling}, the inappropriate use of leaked template can be protected. These different approaches to template protection have varying advantages and limitations and have been studied comprehensively in \cite{patel2015cancelable}. Before designing a template protection scheme, a cost-security analysis is essential and should be encouraged. 

Despite its feasible invertibility, a physical intrinsic feature such as a norm map should be preferred over an extrinsic feature such as BubbleTag because the extrinsic PUFs are \textit{visual} in nature (such as coordinates of bubbles), and inverting them into a plausible synthetic input signal in the digital domain is generally easier than physical features which are a (nearly) \textit{invisible} description of the surface texture. Creating a hashed representation $H(\phi(f(x_{k})))$ also protects against invertibility since even though $x_k$ is recoverable from $\phi(f(x_{k}))$, $H(\phi(f(x_{k})))$ cannot be inverted into $\phi(f(x_{k}))$ for a cryptographically secure $H$.

\subsubsection{Attacks on Stage 4: Decision-Making System}  
\hfill \smallskip \\
The fourth and last stage of the pipeline is the decision-making system that acts as a comparator and matches the test feature with the references present in the database. This is the stage where the final authentication decision is made.

\smallskip
\noindent \textbf{Properties targeted}: \textit{Confidentiality; PUF Uniqueness}

\smallskip
\noindent \textbf{Hill Climbing Attacks}: The choice of the distance metric used to evaluate a match also has security ramifications. There is a large body of work in the field of biometrics that has investigated ``information leaking metrics" \cite{rathgeb2010attacking,muramatsu2008online,spall1998implementation}. If such metrics are used in decision-making logic $\delta$ and the distance scores are observable, information is leaked and hill-climbing attacks \cite{galbally2007bayesian} are possible. Hill-climbing attacks solve an optimization problem of creating a query that will result in successful authentication, given enough observations of the leaking metric \cite{maiorana2014hill}. 

\smallskip
\noindent \textbf{Mitigation Strategies}: Hill climbing attacks can be defended by using a class of distance metrics that does not leak information, or by limiting the number of queries that can be sent to the authentication system via a lockout mechanism. The class of $\ell_P$ norms as a similarity metric has been shown to leak information about raw templates \cite{pagnin2014leakage}, thus adopting other metrics such as Pearson correlation $\rho$ should be preferred.

\subsubsection{Attacks on Interfaces between Modules}  
\hfill \smallskip \\
The interfaces between the different modules are also an important constituent of the attack surface.

\smallskip
\noindent \textbf{Properties targeted}: \textit{Replay Resistance; Confidentiality; Integrity}

\smallskip
\noindent \textbf{Replay Attacks}: There can be man-in-the-middle attacks, where the adversary can intercept the estimated features for the counterfeit product and replace it with some observed registered norm map. The adversary can keep using the same observed features. These attacks are often used in conjunction with attacks on other stages such as hill-climbing or template leakage. 

\noindent \textbf{Mitigation Strategies}: A simple way to deal with such an attack is to use strong cryptographic protocols such as TLS \cite{dierks1999tls} and common security primitives such as time-stamping \cite{lam1992freshness}.

%% file: Sections/sec7_practical_sys_design.tex
\section{Practical Anti-Counterfeiting Systems\label{Sec7}}

We consolidate our attack mitigation strategies into a set of best practices for supply chains and then examine real-world application examples by considering supply chains of different scales. We consider authentication systems based on norm maps due to their numerous advantages outlined in the previous sections.

\subsection{Security Best Practices for Supply Chain Authentication Infrastructure\label{bestprac}}

\begin{itemize}
    \item \textbf{Template Protection}: Similar to biometric systems, protecting reference templates counters the most severe adversarial attacks.  
    \item \textbf{Immutable Audit Trails}: The use of comprehensive and non-repudiable audit trails secures against insider attacks and adds transparency to the supply chain.
    \item \textbf{Using Physical Intrinsic Identifiers}: Physical intrinsic identifiers add a layer of obscurity to the system that is only unveiled by domain expertise.
    \item \textbf{Signal Forensics}: Using signal forensics can protect against synthetic inputs, an important strategy due to the power of machine learning.
    \item \textbf{Distributed Infrastructure}: Distributed storage solutions such as blockchain networks ensure availability. This is extremely important due to the risk of large monetary losses. 
    \item \textbf{Cryptographic Security}: Strong cryptographic protocols such as TLS should always be used to secure communication between modules. 
\end{itemize}

\subsection{Small-Scale Supply Chains}

\begin{figure}
    \centering
    \includegraphics[width=0.85\linewidth]{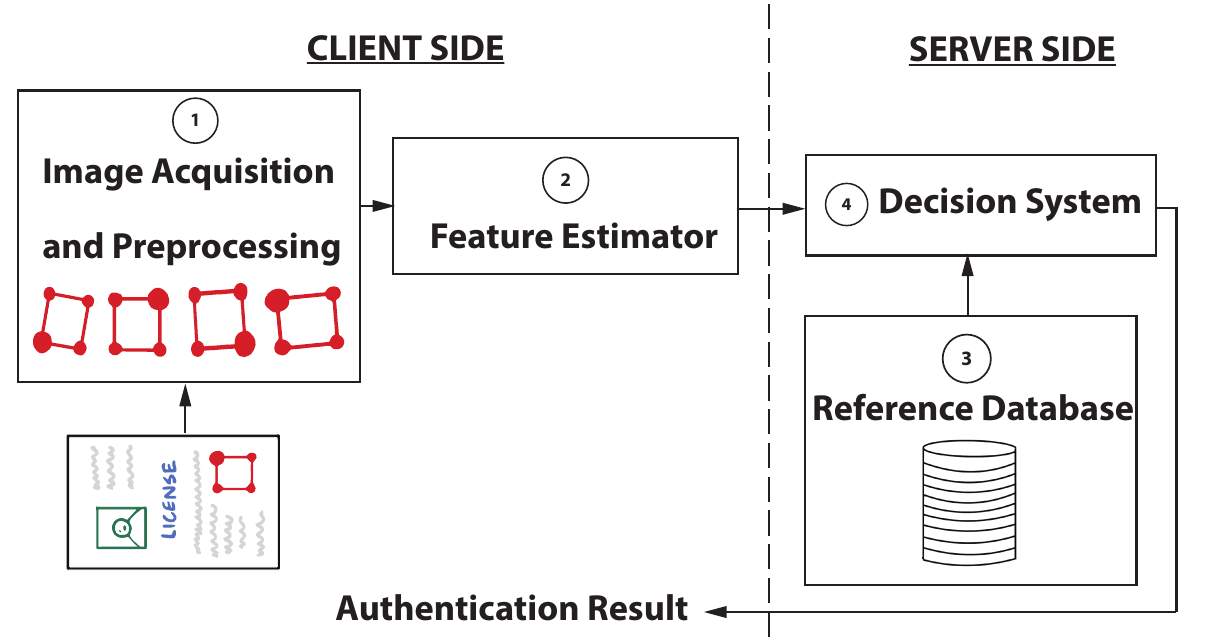}
    \caption{A client--server authentication system}
    \label{fig:clientserver}
\end{figure}

\begin{tcolorbox}[colback=green!5,colframe=green!35!black]
\textbf{Application 1: Art Auction House.} Bob, an artist, approaches the auction house to sell his newest artwork. The auction house accepts the task. He seals his artwork in an envelope and registers it using the paper-PUF authentication system set up by the auction house. Later, Alice wins the bidding process and uses the authentication system to scan and authenticate the envelope containing artwork.
\end{tcolorbox}

\smallskip
\noindent \textbf{Assumptions}: \textit{The supply chain is small and almost all stakeholders are trustworthy.}

\smallskip
\noindent \textbf{Application Scenario}: A client--server model, as illustrated in Fig.~\ref{fig:clientserver}, is an appropriate choice for a supply chain that almost entirely consists of trusted stakeholders. This mutual trust implies that different stakeholders can collaborate to share information and set up a trusted service provider. Additionally, due to the small size of the supply chain, all stakeholders can be properly vetted. In Application 1, Bob can share the envelope with the auction house that can register it after estimating its norm map. The norm-map-based authentication is non-invasive and does not damage the art, making it well-suited for this application. The artwork can be stored while the bidding is done. When Alice wins the bid, the norm map can be estimated again and compared to the reference, demonstrating the authenticity of the envelope containing artwork.

\smallskip
\noindent \textbf{Design Considerations}: The security needs of this application scenario can be supported by deploying a small, private blockchain as the server. This ensures non-repudiability for audits and protects against potential tampering. Since the number of participants is low, vetting the stakeholders is possible. Thorough vetting makes Sybil attacks unlikely because it prevents the creation of a large number of fake identities. Consequently, lightweight consensus mechanisms that usually fall prey to Sybil attacks, are feasible for deployment. The round-robin consensus \cite{xiao2020survey} can be a good alternative design choice to the computationally expensive proof-of-stake \cite{wood2014ethereum} (or proof-of-work \cite{nakamoto2008bitcoin}) consensus. In Application 1, since the auction house is a neutral trusted party for the transaction between two customers, it can do the vetting and take on the role of a service provider by maintaining this blockchain infrastructure.

\smallskip
\noindent \textbf{Security Considerations}: Using a trusted service provider, it is quite straightforward to provide security to the system. If the stakeholders only have access to the image acquisition module and the integrity of the rest of the client is guaranteed, a malicious client cannot circumvent the system. Thus, the best practice in this application scenario is to use a distributed infrastructure such as a blockchain storage network to prevent external hackers from disrupting the system. Thanks to the trusted server, insider threats are not critical. A system designer can and should consider the other best practices mentioned in Section~\ref{bestprac} to add additional layers of protection but maintaining a trusted server is a priority.  

\smallskip
\noindent \textbf{Challenges}: Small-scale supply chains with a trusted centralized service provider represent a scenario in which there is a high amount of established trust between the stakeholders prior to the authentication. Often, supply chains span different geographical territories and are complex. In such a scenario, establishing trust prior to authentication might not be feasible. Furthermore, relying on a centralized service provider can add overhead since there is a considerable cost associated with the setup and maintenance of the service provider's proprietary infrastructure. Analyzing the economics of deploying such a system is an important design consideration.    

\begin{tcolorbox}[colback=green!5,colframe=green!35!black]
\textbf{Application 2: Large E-Commerce Website.} Bob is a sneaker retailer who sources his products from a large e-commerce website. Alice comes to Bob's shop to buy expensive sneakers stored in a sealed cardboard box. Alice uses an open-source app to scan and authenticate the box containing sneakers. 
\end{tcolorbox}

\subsection{Large-Scale Supply Chains}

\smallskip
\noindent \textbf{Assumptions}: \textit{The supply chain is extremely large and only a majority of stakeholders are trustworthy.}

\smallskip
\noindent \textbf{Application Scenario}: In large-scale applications such as in the case of consumer goods, a client--server model can fail to be secure due to the sheer scale of these multi-national supply chains. In such scenarios, it is not easy to choose or establish a centralized authority that everybody trusts. A stakeholder may not know if the other stakeholders are honest or malicious in nature. The only rather weak assumption that one might make is that the majority of participants in these supply chains are honest and want the supply chain to operate successfully.

\smallskip
\noindent \textbf{Design Considerations}: We consider a system model that represents the worst-case design that can incorporate all these practical aspects: a peer-to-peer (P2P) model where each stakeholder is a peer node. We assume that each node or peer in this P2P model has access to all the operational stages of the authentication pipeline 1-4 and the authentication process is carried out locally at each node. Each peer acquires images, preprocesses them, estimates the norm map, and then matches the estimated norm map to the reference norm maps using a decided decision logic. This is a fully decentralized process with no single trusted authority.

\begin{figure} 
    \centering
        \subfloat[\label{fig:7a}]{\raisebox{0.5cm}{
        \includegraphics[width=0.6\linewidth]{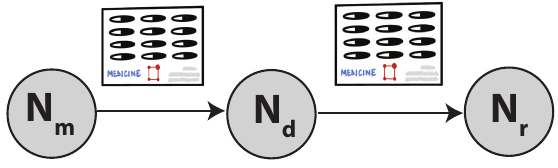}}}
        \subfloat[\label{fig:7b}]{
        \includegraphics[width=0.4\linewidth]{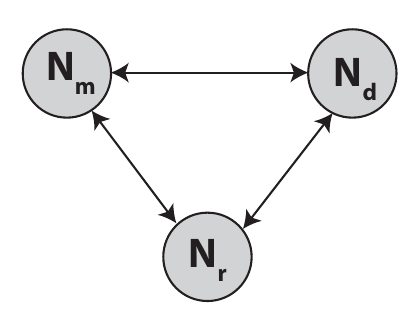}}
    \caption{(a) The product is exchanged between different stakeholders of the supply chain in the physical world. (b) During mutual authentication, stakeholders communicate with each other to verify local authentication at each node. $N_{\text{m}},\ N_{\text{d}}, \ N_{\text{r}}$ are the manufacturer, distributor, and retailer respectively.}
  \label{fig:p2p}
\end{figure}

\smallskip
\noindent \textbf{Security Considerations}: Since every stakeholder/node has \textit{perfect knowledge} and nodes could be malicious, no system integrity can be guaranteed and all the threats discussed in Section~\ref{attackcommon} are possible.

\smallskip
\noindent \textbf{Secure System Design}: Since all attacks are simultaneously possible, one of the few available design choices to secure this authentication process is to introduce regular communication between different nodes and use template protection. To prototype a secure system in the presence of \textit{any} possible attack, we use the assumption that the majority of stakeholders are honest. To ensure non-repudiation of any communication between nodes, we design each node to be a peer of the blockchain network. Each node has a replica of the complete blockchain locally, communicates information about its local authentication with other nodes during the consensus steps, and shares the common blockchain storage that contains \textit{protected reference features}. 

\smallskip
\noindent \textbf{Mutual Authentication:} Doing local authentication and communicating the result is not enough. Since all attacks are feasible simultaneously, an adversary can launch a union of threats and circumvent authentication. Therefore, the nodes need to communicate additional information. Each node must act as a \textit{prover} to other nodes that their local authentication process was legitimate. A simple implementation of this process is shown in Algorithm~\ref{alg:mutualauth}, where each time a prover node does local authentication, they also share their (Image, Hashed norm map) pair with all other nodes. These \textit{verifier} nodes then estimate the hashed norm map from the shared images using their local feature extractor and verify that the hashes match. This provides security since a node cannot invert the hashed references in the local storage into images that pass mutual authentication. 

\RestyleAlgo{ruled}

\begin{algorithm}[t!]

\caption{Mutual Authentication Process}\label{alg:mutualauth}
\SetKwInput{KwAssume}{Assumption}
\SetKwInput{KwResult}{Costello}
\KwAssume{Prover node : $N_{\text{d}}$}
\KwAssume{Verifier nodes : $N_i, \ i \ \in \{1,\dots,N\}$}
\KwAssume{Cryptographically secure hash : $H$}
\SetKwRepeat{Do}{do}{while}
\While{Consensus Step == True}{
    $\forall \ i,\ N_{i} \gets \ (x_q,H(\phi(f(x_{\text{q}})))) \ \text{from} \ N_{\text{d}}$;
    
    $B \gets \ \mathds{1}(H(\phi(f(x_{q}))) = H(\phi(f(x_{K}))))$;
    
    \eIf{$B == 1$}{
    $\forall \ i \ \in \ Consensus \ Set,\ N_{i}$ signs off on $N_{\text{d}}$'s verification;
    }{\If{$B == 0 $}
    {Flag $N_{\text{d}}$ as malicious;}
    }
}

\end{algorithm}

\smallskip
\noindent \textbf{Challenges}: Algorithm \ref{alg:mutualauth} has limitations. Even though only one node undergoes the authentication process for a product $k$, all $N$ nodes need to perform norm map estimation. This introduces a significant cost overhead. Providing every other node with the (Image, Hashed norm Map) pair leads to the leakage of genuine pairs to malicious nodes. In a future mutual authentication cycle, these malicious nodes can use this information to circumvent authentication. Mutually authenticating products also requires high bandwidth since the pairs are shared with all $N$ nodes during each consensus cycle. 

\smallskip
\noindent \textbf{Improvements}: Algorithm~\ref{alg:mutualauth} can be improved. For example, the mutual authentication algorithm can be run every $M$ consensus cycle. This lowers overhead but implies that $M$ malicious nodes can be missed in the worst case, hence making the identification of malicious nodes ambiguous. For combating information leakage, previously used pairs (and a close perturbation space) should be checked. It should be ensured that creating the hashed norm map via brute force is expensive and infeasible. Finally, to protect against local threats, all best practices in Section~\ref{bestprac} are necessarily needed at each node, and the results should be shared with verifiers if bandwidth permits. 

\subsection{Medium-Scale Supply Chains}

\begin{tcolorbox}[colback=green!5,colframe=green!35!black]
\textbf{Application 3: Semiconductor Companies.} Bob owns a semiconductor company that makes printed circuit boards (PCBs). Alice's company wants to buy these PCBs. Bob seals the PCBs in a cardboard box, registers them to a paper-PUF authentication system, and sells these PCBs to Alice. She uses Bob's infrastructure to authenticate them and ships the PCBs to a foreign contractor, who scans the box via an open-source app to perform authentication.
\end{tcolorbox}

\smallskip
\noindent \textbf{Assumptions}: \textit{This supply chain's assumptions fall between the large-scale and small-scale scenarios.}

\smallskip
\noindent \textbf{Application Scenario}: Bob and Alice are located in different countries that have trade agreements. However, Alice is outsourcing her chip assembly to a factory that is located in a region without any such agreements with either Alice or Bob's country. 

\smallskip
\noindent \textbf{Design Considerations}: For medium-scale business-to-business (B2B) business models such as companies in the semiconductor industry, hybrid models are the most appropriate choice since they balance both scale and security. In this case, Alice and Bob can set up a server infrastructure for the part of the supply chain in her country while the P2P model operates in the untrustworthy region. 

\smallskip
\noindent \textbf{Security Considerations}: The security implications are determined by the evaluation of the client--server and P2P models since they represent the two extremes of the trust spectrum.

%% file: Sections/sec8_common_sysdesign.tex
\section{Common Design Considerations\label{Sec8}}

Finally, we examine some common design choices related to the systems across all three scales.

\smallskip
\noindent \textbf{Implementation of Reference Database $\mathcal{D}$}: To implement large-scale approximate-nearest-neighbor (ANN) based storage solutions, system designers can utilize space partitioning data structures, for example, the \textit{kd}-tree \cite{kdtree}. However, \textit{kd}-tree suffers from the curse of dimensionality \cite{10.1145/3292500.3330875}. Since norm maps have dimensionality akin to images, such variants as quadtree \cite{samet1984quadtree}, octree \cite{MEAGHER1982129}, ball trees \cite{omohundro1989five}, and locality-sensitive hashing (LSH) \cite{gionis1999similarity} can be a more scalable design choice.

\smallskip
\noindent \textbf{Storage and Scalability}: There is a strong need to evaluate the cost of the infrastructure associated with choosing a particular storage solution and understand the trade-off of cost and security. For tamper-proof solutions such as blockchains, the amount of data stored on the blockchain grows drastically with time due to their immutable nature. To store data with blockchains more efficiently, optimizations have been proposed \cite{heo2024blockchain}; replication-based solutions \cite{kumar2019implementation,zheng2018innovative,xu2018cub,dai2019jidar,heo2022blockchain}, redaction-based solutions \cite{pyoung2019blockchain}, and content-based solutions \cite{arslan2022compress}. Since most paper-based authentication systems use multi-dimensional features, the storage requirements of $\mathcal{D}$ need to be explicitly considered for practical deployability. If one is leveraging permissionless blockchains, for instance, Ethereum \cite{wood2014ethereum} for implementing $\mathcal{D}$, gas costs under this blockchain framework can help quantitatively analyze the cost-security trade-off. 

\smallskip
\noindent \textbf{Revealing Partial Information}: To add another layer of security to these systems, the creation of a hold-out reference set should be explored. For example, based on the multi-resolution principle in visual processing \cite{vaidyanathan_1993}, this hold-out set would store a \textit{finer} representation of the templates that can be used to refine the authentication confidence from the normal verification process which uses \textit{coarse} representations. A template can be decomposed into sub-templates containing \textit{finer} and \textit{coarser} representations by signal processing techniques such as sub-band decomposition \cite{vaidyanathan_1993}. Existing works have not explored such ideas in detail, but work by Liu et al. \cite{flatbed} that creates 3D height maps of norm maps based on a difference-of-Gaussian (DoG) representation \cite{lindeberg1994scale,lowe2004distinctive} shows promise.

%% file: Sections/sec9_discussion.tex
\section{Discussions\label{Sec9}}

\noindent
\textbf{Related SoK/Surveys:} To our knowledge, no other survey or SoK comprehensively covers paper-PUF-based anti-counterfeiting systems. Surveys often look at the security of hardware PUFs for cryptographic protocols \cite{shamsoshoara2020survey,ruhrmair2013pufs,dey2021puf} or at specific types of techniques used to create PUFs \cite{zerrouki2022survey}. Different from these works, we take a holistic view of anti-counterfeiting systems, create a framework that evaluates different prior works, explore multiple systems based on area of application, and perform a systematic security analysis. Some prior works consider the interactions of the physical and cyber world explicitly. The authors in \cite{aniello2019towards,islam2022integrating} developed systems consisting of proprietary PUF-based tags which are linked to a blockchain, and the work of \cite{flatbed} briefly proposed practical norm-map based systems with some threat modeling. 

\smallskip
\noindent
\textbf{Concluding Thoughts:} As the demand for physical goods continues to grow in scale, there is an urgent need to develop robust solutions to combat counterfeits in order to prevent harm to public safety and health. Our work investigates the supply chain ecosystem in its entirety by understanding the interactions of the two worlds that compose it. By investigating key practical systems based on plausible real-world scenarios, we find that there is no one all-encompassing solution. Rather, different application scenarios require varying system designs. Future anti-counterfeiting solutions can leverage our analysis as a starting point to formulate practical and secure systems.

%% file: Sections/acknowledgments.tex
\section*{Acknowledgments}

This work is supported by the US National Science Foundation (award numbers ECCS-2227261 and ECCS-2227499).